\documentclass[submission,copyright,creativecommons]{eptcs}

\providecommand{\event}{iFMCloud 2016} 

\usepackage{epsf}
\usepackage{listings}
\usepackage{array}
\usepackage{longtable}
\usepackage{multirow}
\usepackage{mathrsfs}
\usepackage{amsmath, amsthm, amssymb, wasysym}
\usepackage[all]{xy}
\usepackage{rotating}

\usepackage{xspace}
\usepackage{framed}
\usepackage[framemethod=TikZ]{mdframed}
\usepackage{wrapfig}
\usepackage{pgf}
\usepackage{tikz}
\usepackage{reotex}
\usepackage{setspace}
\usepackage{caption}
\usepackage{rotating}

\def\ie{\emph{i.e.}}

\def\reo{\textsf{Reo}\xspace}

\def\sreo{\textsf{Stochastic Reo}\xspace}

\def\imcreo{\ensuremath{\sf IMC_{Reo}}\xspace}
\def\dimcreo{\ensuremath{\mathcal{D}\sf{IMC_{Reo}}}\xspace}

\def\nodeset{\ensuremath{\mathcal{N}}\xspace}

\def\imcbufset{\ensuremath{\mathfrak{Q}}\xspace}

\def\syncname{sync}
\def\lossyname{lossy}
\def\drainname{drain}
\def\fifoname{fifo}

\def\fifoename{fifo_e}

\def\replicatornodename{replicator}
\def\mergernodename{merger}
\def\routernodename{router}
\def\mrname{merger\!\!-\!\!replicator}
\def\mxname{merger\!\!-\!\!router}

\newcommand{\reochannel}[1]{\ensuremath{\mathsf{#1}}\xspace}

\newcolumntype{C}[1]{>{\centering\arraybackslash}m{#1\textwidth}}

\newcommand{\partsof}[1]{\ensuremath{2^{#1}}} 

\newcommand{\enset}[2][\normalsize]{{#1 \ensuremath{ \{ #2 \} }}}

\newcommand{\setdef}[2]{\{#1|\;#2\}}   

\newcommand{\realsetpos}{\ensuremath{\mathbb{R}^+}\xspace} 

\newcommand{\sse}{\; \mathsf{iff} \;} 

\newcommand{\bisim}{\sim} 

 % ...

\makeatletter
\newcommand{\superimpose}[2]{%
  {\ooalign{$#1\@firstoftwo#2$\cr\hfil$#1\@secondoftwo#2$\hfil\cr}}}
\makeatother

\def\node#1{\ensuremath{\underline{#1}}}

\newcommand{\enbox}[2][\tiny]{{#1\tikz[baseline=-0.25\baselineskip]{\draw node[rectangle, minimum height=0cm,minimum width=0cm, inner ysep=1.5pt, inner xsep=1.5pt, draw] {\ensuremath{#2}};}}}

\newcommand{\entrans}[2][\scriptsize]{{#1 \tikz[baseline=-0.25\baselineskip]{\draw node[rectangle, minimum height=0cm,minimum width=0cm, inner sep=-1pt] {\ensuremath{\{#2\}}};}}}

\newcommand{\encircle}[3][0mm]{
\node[circle, dashed, fit=(#3), inner sep=#1, draw] (#2) {} ; }

\newcommand{\enqueue}[2][\tiny]{{#1\tikz[baseline=-0.25\baselineskip]{\draw node[ellipse, minimum height=0cm,minimum width=0cm, inner ysep=0.2pt, inner xsep=0.5pt,draw] {\ensuremath{#2}};}}}

\newcommand{\dequeue}[2][\tiny]{{#1\tikz[baseline=-0.25\baselineskip]{\draw node[ellipse, minimum height=0cm,minimum width=0cm, inner ysep=0.2pt, inner xsep=0.5pt,fill={gray!50!black}, text={white}, draw] {\ensuremath{#2}};}}}

\newcommand{\intertrans}[1][0.6]{\tikz[baseline=-0.25\baselineskip]{
\draw[->, dashed] (0,0) -- (#1,0);
}}

\newcommand{\marktrans}[1][0.6]{\tikz[baseline=-0.25\baselineskip]{
\draw[->] (0,0) -- (#1,0);
}}

\def\restricts{\!\!\restriction}

\def\nactblock{\ensuremath{\ntriangleright}}

\newcommand{\composevia}[1][M]{\ensuremath{\ \parallel_#1\ }}

\def\1{202}
\def\2{203}
\def\3{204}
\def\4{205}

\usetikzlibrary{arrows,%
                automata,%
                backgrounds,%
                decorations.markings,%
                decorations.shapes,%
                decorations.text,%
                shapes.geometric,%
                shapes.misc,%
                shapes.symbols,%
                shapes.multipart,%
                calc,%
                fit,%
                matrix,%
                positioning}

\newtheorem{definition}{Definition}
\newtheorem{theorem}{Theorem}

\title{An Enhanced Model for Stochastic Coordination}
\author{Nuno Oliveira
\institute{HASLab - INESC TEC\\ Universidade do Minho, Braga, Portugal}
\email{\quad nuno43549@gmail.com}
\and
Luis Soares Barbosa
\institute{HASLab - INESC TEC\\
Universidade do Minho, Braga, Portugal \thanks{Luis S. Barbosa is supported by grant \texttt{SFRH/BSAB/113890/2015} from FCT, the Portuguese Foundation for Science and Tecnhology. This research is financed by the ERDF
COMPETE 2020 Programme  
within project  \texttt{POCI-01-0145-FEDER-00696}, and by National Funds through FCT as part 
of project  UID/EEA/50014/2013.}}
\email{ lsb@di.uminho.pt}
}
\def\titlerunning{An Enhanced Model for Stochastic Coordination}
\def\authorrunning{N. Oliveira \& L. S. Barbosa}
\begin{document}
\maketitle

\begin{abstract}
Applications developed over the cloud coordinate several, often anonymous, computational resources, distributed over different execution nodes, within flexible architectures. Coordination models able to represent quantitative data provide a powerful basis for their analysis and validation. This paper extends \imcreo, a semantic model for \sreo based on interactive Markov chains, to enhance its scalability, by regarding each channel and node, as well as interface components, as independent stochastic processes that may (or may not) synchronise with the rest of the coordination circuit.
\end{abstract}

\section{Introduction}
The increasing ubiquity and complexity of  cloud  applications
and their management brings research on coordination languages and models \cite{GC92}  up front as a main tool for design and analysis. On the one hand, this opens an interesting 
opportunity for formal methods; on the other it clearly challenges their scalability.

This paper  addresses such a challenge from a specific stand point: that of the \reo coordination model \cite{arbab2002,arbab04} and its stochastic version \cite{arbab2009,moon2011}. In a previous paper \cite{OSB15}, the authors, in collaboration with Alexandra  Silva, proposed a semantic model  for \sreo based on interactive Markov chains \cite{hermanns2002}.  
The model, known as \imcreo, is compositional and has the advantage of bringing to the coordination community a panoply of tools developed for quantitative analysis of probabilistic transition systems. Due to a  rapid state explosion, the (use of the) model, however,  does not scale up to the point of being really useful for analysis of big coordination scenarios, as found in typical cloud applications.

The paper starts in Section 2 with a brief review of \reo, its stochastic version, and  \imcreo. Due to space restrictions such introductions are necessarily very short; the interested reader is referred to the relevant literature \cite{moon2014,OSB15} for details. Sections 3 and 4 introduce an enhanced model which smoothly extends \imcreo, increasing its ability to deal with bigger and more complex coordination protocols in a stochastic setting. The proposed model, called \dimcreo, from \emph{distilled} \imcreo, relaxes the basic \reo  assumption on mixed nodes as self-pumping stations \cite{arbab04}, which allowsfor data to be read and written with no processing delay. In practice, namely for cloud based applications,  this assumption is unrealistic: I/O operations take time and, therefore, may interfere with QoS values. Finally, section 5 concludes.

\section{Background}
\paragraph{\reo.}

\reo~\cite{arbab2002,arbab04} is a channel-based model for the exogenous coordination of components in the context of component-based software. 
A channel is a directed communication mean with exactly two ends: a source and a sink end; but \reo also accepts undirected channels (\ie\ channels with two ends of the same sort). 
A channel is synchronous when it delays the operations at each of its ends so that they can succeed simultaneously. Otherwise it is  asynchronous, exhibiting memory capabilities or the possibility of specifying an ordering policy for content delivery. Moreover, a channel may also be lossy when it delivers some values but loses others depending on a specified policy. Figure~\ref{fig:stoch_reo_channels} recalls the basic  channels used in \reo, represented, however, in their stochastic version.
The \reochannel{\syncname} channel transmits data from one end to another whenever there is a request at both ends synchronously, otherwise one request shall wait for the other. 
The \reochannel{\lossyname} channel behaves likewise, but data may be lost whenever a request at the source end is not matched by another one at the sink end. Differently, a \reochannel{\fifoname} channel has buffering capacity of (usually) one memory position, therefore
allowing for asynchronous occurrence of input/output requests.  The qualifiers $\mathsf{e}$ or $\mathsf{f}$ refer to the channel internal state (either
\emph{empty} or \emph{full}).
Finally, the \reochannel{\drainname} channel accepts data synchronously at both ends and  loses it.

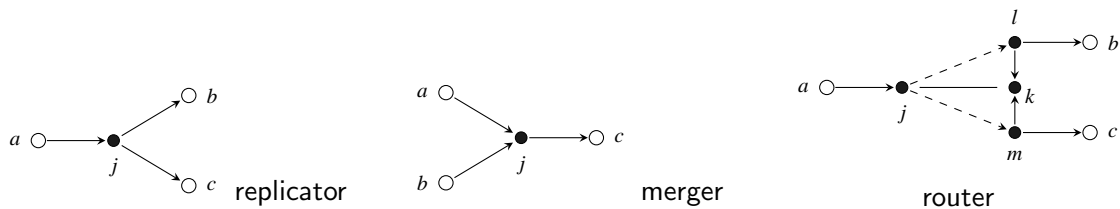
\begin{figure}[htbp]
\begin{center}
\begin{minipage}[b]{0.32\linewidth}
\begin{center}

\begin{tikzpicture}
\ionode{(in)}{(0,0)}{node[left, xshift=-3pt]{\scriptsize $a$}}
\ionode{(out1)}{(2,0.6)}{node[right, xshift=3pt]{\scriptsize $b$}}
\ionode{(out2)}{(2,-0.6)}{node[right, xshift=3pt]{\scriptsize $c$}}
\mixednode{(j)}{(1,0)}{node[below, yshift=-3pt]{\scriptsize $j$}}

\sync{(in)}{(j)}{}
\sync{(j)}{(out1)}{}
\sync{(j)}{(out2)}{}

\end{tikzpicture}
\vspace{0.1cm}
\reochannel{replicator}

\end{center}
\end{minipage}
\begin{minipage}[b]{0.32\linewidth}
\begin{center}

\begin{tikzpicture}
\ionode{(in1)}{(0,0.6)}{node[left, xshift=-3pt]{\scriptsize $a$}}
\ionode{(in2)}{(0,-0.6)}{node[left, xshift=-3pt]{\scriptsize $b$}}
\ionode{(out2)}{(2,0)}{node[right, xshift=3pt]{\scriptsize $c$}}
\mixednode{(j)}{(1,0)}{node[below, yshift=-3pt]{\scriptsize $j$}}

\sync{(in1)}{(j)}{}
\sync{(in2)}{(j)}{}
\sync{(j)}{(out2)}{}

\end{tikzpicture}
\vspace{0.1cm}
\reochannel{merger}

\end{center}
\end{minipage}
\begin{minipage}[b]{0.32\linewidth}
\begin{center}

\begin{tikzpicture}
\ionode{(in)}{(0,0)}{node[left, xshift=-3pt]{\scriptsize $a$}}
\ionode{(out1)}{(3.5,0.6)}{node[right, xshift=3pt]{\scriptsize $b$}}
\ionode{(out2)}{(3.5,-0.6)}{node[right, xshift=3pt]{\scriptsize $c$}}

\mixednode{(l)}{(2.5,0.6)}{node[above, yshift=3pt]{\scriptsize $l$}}
\mixednode{(m)}{(2.5,-0.6)}{node[below, yshift=-3pt]{\scriptsize $m$}}
\mixednode{(j)}{(1,0)}{node[below, yshift=-3pt]{\scriptsize $j$}}
\mixednode{(k)}{(2.5,0)}{node[right, yshift=-3pt]{\scriptsize $k$}}

\sync{(in)}{(j)}{}
\lossysync{(j)}{(l)}{}
\lossysync{(j)}{(m)}{}
\syncdrain{(j)}{(k)}{}
\sync{(l)}{(k)}{}
\sync{(m)}{(k)}{}
\sync{(l)}{(out1)}{}
\sync{(m)}{(out2)}{}

\end{tikzpicture}
\vspace{0.1cm}
\reochannel{router}

\end{center}
\end{minipage}

\caption{\reo connectors}
\label{fig:rep_merger_router}
\end{center}
\end{figure}

Channels are composed to define more complex coordination structures referred to as connectors. Composition of channels is made on their ends, giving rise to nodes.
A node may be of three distinct types: 
$(i)$~source node, if it connects only source channel ends; 
$(ii)$~sink node, if it connects only sink channel ends and
$(iii)$~mixed node, if it connects both source and sink channel ends.
The first two types may also be referred to as the connector's ports. Figure~\ref{fig:rep_merger_router} presents three such connectors.

As expected of any compositional model, \reo connectors behaviour arise from the behaviour of each constituent channel. However, as composition is made on channel ends, originating nodes, also these nodes contribute to the overall connector behaviour.
The connectors of Figure~\ref{fig:rep_merger_router} actually encode the simple form of three of these nodes. The \reochannel{replicator} connector replicates data flowing from port $a$ to ports $b$ and $c$, in parallel, through mixed node $j$~---~the replicator node. This behaviour, which is synchronous, only holds when there are pending requests in all the connector ports. The \reochannel{merger} connector merges data coming from ports $a$ and $b$ to port $c$, through mixed node $j$~---~the merger node. The merge of data is synchronous but only on two ends at each time: either on $a$ and $c$ or on $b$ and $c$. This means that node $j$ performs a non-deterministic choice when there are pending requests at all the boundary ports, preventing one of the input ports from firing. The \reochannel{router} connector, usually represented as $\tikz[baseline=-3pt]{\xrouter{(xr)}{(0,0)}{}}$, is a mutual exclusive router of data, taking data from input port $a$ into either port $b$ or port $c$, depending on the existence of pending requests at the output ports. When there are pending requests at the same time in both output ports, mixed node $k$~---~the router~---~non-deterministically choses  (since it encodes a merge) which of the two ports will synchronously fire: either $a$ and $b$ or $a$ and $c$.

\paragraph{\sreo.} \sreo~\cite{arbab2009,moon2011}
extends \reo by modelling coordination from a quantitative perspective. Non-negative real (stochastic) values are added both to channels and to their ends to represent, respectively, \emph{processing delays} and IO \emph{arrival rates}. 
The former models the time needed for the channel to process data from one point to another, where point refers to a channel end, a buffer or a point where data is lost or automatically produced. Each channel, depending on its type, may be annotated with more than one processing delays. 
Arrival rates model the time between consecutive arrivals of environment-issued IO operations to channel ends. 
Figure~\ref{fig:stoch_reo_channels} shows the basic channels of stochastic \reo, represented as normal \reo channels, but annotated with stochastic values (rates and delays). 

\begin{figure}[htbp]
\begin{center}
\begin{minipage}[b]{0.23\linewidth}
\begin{center}

\tikz \sync{(0,0)}{(1.5,0)}{
	node[above]{$\gamma_{ab}$} 
	node[below,xshift=20pt]{$\gamma_b$} 
	node[below,xshift=-20pt]{$\gamma_a$}
};
\vspace{0.1cm}

\reochannel{\syncname}
\end{center}
\end{minipage}
\begin{minipage}[b]{0.23\linewidth}
\begin{center}
\tikz \lossysync{(0,0)}{(1.5,0)}{
	node[above]{$\gamma_{ab}$} 
	node[below]{$\gamma_{aL}$} 
	node[below,xshift=20pt]{$\gamma_b$} 
	node[below,xshift=-20pt]{$\gamma_a$}
} ;
\vspace{0.1cm}

\reochannel{\lossyname}
\end{center}
\end{minipage}
\begin{minipage}[b]{0.23\linewidth}
\begin{center}
\tikz \syncdrain{(0,0)}{(1.5,0)}{
	node[above]{$\gamma_{ab}$} 
	node[below,xshift=20pt]{$\gamma_b$} 
	node[below,xshift=-20pt]{$\gamma_a$}
} ; 
\vspace{0.1cm}

\reochannel{\drainname}
\end{center}
\end{minipage}
\begin{minipage}[b]{0.23\linewidth}
\begin{center}
\tikz \fifoe{(0,0)}{(2.5,0)}{
	node[above, xshift=-45pt]{$\gamma_{aB}$} 
	node[above]{$\gamma_{Bb}$} 
	node[below,xshift=10pt]{$\gamma_b$} 
	node[below,xshift=-55pt]{$\gamma_a$}
} ;
\reochannel{\fifoename}
\end{center}
\end{minipage}

\caption{Primitive \sreo channels.}
\label{fig:stoch_reo_channels}
\end{center}
\end{figure}

Stochastic \reo is still compositional. Processing delays of each individual channel in a composition scenario are not changed. The request arrival rates, however, are only preserved for the boundary nodes of the connector. As mixed nodes are internal (hidden from the exterior) the arrival request rates associated to the constituent channel ends are ignored, which means that these nodes are always ready to read/write data from/to the channels. This behaviour is known as the {\em self-contained pumping station}, firstly referred in~\cite{arbab04}.

\paragraph{$\imcreo$.}\label{sec:imcreo}

In a previous paper \cite{OSB15}, the authors introduced a compositional semantic model for \sreo\, based on interactive Markov chains 
\cite{hermanns2002}, a formalism combining continuous-time Markov chains\cite{baier2003,aziz2000} with
process algebra~\cite{baeten2005}.
The model is state-based, states capturing the possible behaviour of a connector: data arrivals and data flowing through ports. Consider sets \nodeset and \imcbufset  of port names and  internal state names, respectively.  
Each state in \imcreo is a triple $(R,T,Q)$, where 
$R, T\in \partsof{\nodeset}$ denote sets of ports/nodes with, respectively, pending requests and data being transmitted; and 
$Q\in \imcbufset$ is an internal state identifier. The latter is used  to distinguish between control states in state-based connectors. For example, in a \reochannel{\fifoname} channel it may indicate whether the buffer is empty or full, by taking $\imcbufset = \enset{\mathsf{empty}, \mathsf{full}}$. 
Markovian transitions are labelled by $\gamma\in \mathbb{R}^+$. Distribution parameter $\gamma$ encodes, in each case, the connector processing delays  and the rates of data arrival at its ports.
Interactive transitions, on the other hand, are labelled with a set $F$ of ports which, on firing, allow data to flow through them. Such ports correspond to the set of actions observable at the relevant \imcreo state. In the sequel, this set is referred to as \emph{actions}, for simplicity. 
The decision to take sets of actions (rather than a single action) to label interactive transitions was crucial to correctly capture (atomic) synchrony in the semantics of \reo. 
In fact, ports firing synchronously to enable data flow are the rule rather than the exception in \reo.
Formally,

\begin{definition}\label{def:imcreo} An \imcreo model is a tuple (S, Act, \intertrans, \marktrans, s), where $S \subseteq Act \times Act\times \imcbufset$ is a nonempty set of states;
$Act \subseteq \partsof{\nodeset}$ is a set of actions (the alphabet); 
$\intertrans \subseteq S \times Act \times S$ is the interactive transition relation; 
$\marktrans \subseteq S \times \realsetpos \times S$ is the Markovian transition relation; and $s \in S$ is the initial state.
\end{definition}

\noindent
Markovian transitions $(s, \gamma, s^\prime)$ are written as $s \stackrel{\gamma}{\marktrans} s^\prime$; whereas 
notation $s \stackrel{a_1a_2...}{\intertrans[1]} s^\prime$ is used for 
interactive transitions $(s, \enset{a_1,a_2,...}, s^\prime)$.
An interactive transition with an empty set of actions is said to be unobservable and is denoted by $s \stackrel{\tau}{\intertrans} s^\prime$.
States of the form $(R,\emptyset,Q)$ are referred to as \emph{request} states and depicted as \enbox[\scriptsize]{R}$_Q$; 
states of the form $(\emptyset, T, Q)$ are referred to as \emph{transmission} states and depicted as \enset{T}$_Q$; states of the form $(R,T,Q)$ are called \emph{mixed} states and are depicted as \enbox[\scriptsize]{R}\enset{T}$_Q$; finally, states of the form $(\emptyset,\emptyset, Q)$ are represented as $\emptyset_Q$ and denote the absence of both requests and data transmissions.
For all representations, the buffer qualifier $Q$ may be omitted, whenever clear from the context.

\begin{figure}[!h]
\begin{center}

\begin{minipage}{0.45\linewidth}
\begin{center}
\tikz[>=latex, shorten >=1pt, node distance=1.5cm, style={font=\scriptsize}, initial where={above}, initial text=]{%
\node[initial] (e) {$\emptyset$}; %
\node (ar) [right of = e] {\enbox[\tiny]{a}} ; %
\node (br) [below of = e] {\enbox[\tiny]{b}} ; %
\node (abr) [below of = ar] {\enbox[\tiny]{a,b}} ; %
\node (abf) [right of = abr] {\entrans{a,b}} ; %
\path[->] (e) edge node[above]{$\gamma_a$} (ar); 
\path[->] (e) edge node[left] {$\gamma_b$} (br); %
\path[->] (ar) edge node[left]{$\gamma_b$} (abr); %
\path[->] (br) edge node[above]{$\gamma_a$} (abr); %
\path[->, dashed] (abr) edge node[above]{$ab$} (abf); %
\path[->] (abf.90) edge[out=130, in=340] node[xshift=24pt]{$\gamma_{ab}$} (e.340); 
} 
\\[0.3cm]
\tikz \sync{(0,0)}{(1.5,0)}{
	node[above]{$\gamma_{ab}$} 
	node[below,xshift=20pt]{$\gamma_b$} 
	node[below,xshift=-20pt]{$\gamma_a$}
};
\tikz \syncdrain{(0,0)}{(1.5,0)}{
	node[above]{$\gamma_{ab}$} 
	node[below,xshift=20pt]{$\gamma_b$} 
	node[below,xshift=-20pt]{$\gamma_a$}
} ; 
\\[0.3cm]
\reochannel{\syncname} and \reochannel{\drainname}
\end{center}
\end{minipage}
\begin{minipage}{0.45\linewidth}
\begin{center}
\tikz[>=latex, shorten >=1pt, node distance=1.5cm, style={font=\scriptsize}, initial where={above}, initial text=]{ 
\node[initial] (e) {$\emptyset$}; 
\node (ar) [right of = e] {\enbox[\tiny]{a}} ; 
\node (br) [below of = e] {\enbox[\tiny]{b}} ; 
\node (abr) [below of = ar] {\enbox[\tiny]{a,b}} ; 
\node (abf) [right of = abr] {\entrans{a,b}} ; 
\node (af) [right of = ar] {\entrans{a}} ; 
\path[->] (e) edge node[above]{$\gamma_a$} (ar); 
\path[->] (e) edge node[left] {$\gamma_b$} (br); 
\path[->] (ar) edge node[left]{$\gamma_b$} (abr); 
\path[->] (br) edge node[above]{$\gamma_a$} (abr); 
\path[->, dashed] (abr) edge node[above]{$ab$} (abf); 
\path[->] (abf.90) edge[out=130, in=340] node[xshift=24pt]{$\gamma_{ab}$} (e.340); 
\path[->, dashed] (ar) edge node[above]{$a$} (af);
\path[->] (af.110) edge[bend right] node[above]{$\gamma_{aL}$} (e.60);
}
\\[0.3cm]
\tikz \lossysync{(0,0)}{(1.5,0)}{
	node[above]{$\gamma_{ab}$} 
	node[below]{$\gamma_{aL}$} 
	node[below,xshift=20pt]{$\gamma_b$} 
	node[below,xshift=-20pt]{$\gamma_a$}
} ;
\\[0.3cm]
\reochannel{\lossyname}
\end{center}
\end{minipage}

\begin{center}
\rule{0.7\linewidth}{0.5pt}\\[0.1cm]
\begin{tikzpicture}[>=latex, shorten >=1pt, node distance=1.5cm, style={font=\scriptsize}, initial where={left}, initial text=]
\node[initial] (e) {$\emptyset_e$}; 
\node (ar) [right of = e] {\enbox[\tiny]{a}$_e$} ; 
\node (br) [below of = e] {\enbox[\tiny]{b}$_e$} ; 
\node (abr) [below of = ar] {\enbox[\tiny]{a,b}$_e$} ; 
\node (taB) [right of = ar] {\entrans{a}$_e$} ; 
\node (brtaB) [right of = abr] {\enbox[\tiny]{b}\entrans{a}$_e$} ; 
\node (efull) [right of = taB] {$\emptyset_f$} ; 
\node (arfull) [right of = efull] {\enbox[\tiny]{a}$_f$} ; 
\node (brfull) [right of = brtaB] {\enbox[\tiny]{b}$_f$} ; 
\node (abrfull) [right of = brfull] {\enbox[\tiny]{a,b}$_f$} ; 
\node (tBbfull) [below of = brtaB] {\entrans{b}$_f$} ; 
\node (artBbfull) [below of = brfull] {\enbox[\tiny]{a}\entrans{b}$_f$} ; 
\node (_erep) [below of = abr]{$\emptyset_e$}; 
\encircle[-1mm]{erep}{_erep}
\node (_arrep) [below of = abrfull] {\enbox[\tiny]{a}$_e$} ;
\encircle[-1mm]{arrep}{_arrep}
\path[->] (e) edge node[above]{$\gamma_a$} (ar); 
\path[->, dashed] (ar) edge node[above]{$a$} (taB); 
\path[->] (taB) edge node[above]{$\gamma_{aB}$} (efull);
\path[->] (efull) edge node[above]{$\gamma_{a}$} (arfull); 
\path[->] (e) edge node[left] {$\gamma_b$} (br); 
\path[->] (ar) edge node[left]{$\gamma_b$} (abr); 
\path[->] (taB) edge node[left]{$\gamma_b$} (brtaB); 
\path[->] (efull) edge node[left]{$\gamma_b$} (brfull); 
\path[->] (arfull) edge node[left]{$\gamma_b$} (abrfull);
\path[->, dashed] (brfull) edge node[left]{$b$} (tBbfull); 
\path[->] (br) edge node[above]{$\gamma_a$} (abr); 
\path[->, dashed] (abr) edge node[above]{$a$} (brtaB); 
\path[->] (brtaB) edge node[above]{$\gamma_{aB}$} (brfull); 
\path[->] (brfull) edge node[above]{$\gamma_{a}$} (abrfull); 
\path[->, dashed] (abrfull) edge node[right]{$b$} (artBbfull); 
\path[->] (artBbfull) edge node[below]{$\gamma_{Bb}$} (arrep); 
\path[->] (tBbfull) edge node[below]{$\gamma_{Bb}$} (erep); 
\path[->] (tBbfull) edge node[below]{$\gamma_{a}$} (artBbfull); 

\end{tikzpicture}
\\[0.3cm]
\tikz \fifoe{(0,0)}{(2.5,0)}{
	node[above, xshift=-45pt]{$\gamma_{aB}$} 
	node[above]{$\gamma_{Bb}$} 
	node[below,xshift=10pt]{$\gamma_b$} 
	node[below,xshift=-55pt]{$\gamma_a$}
} ;
\\[0.3cm]
\reochannel{\fifoename}
\end{center}

\caption{IMC for the basic stochastic \reo channels.}
\label{fig:imcreochannels}
\end{center}
\end{figure}
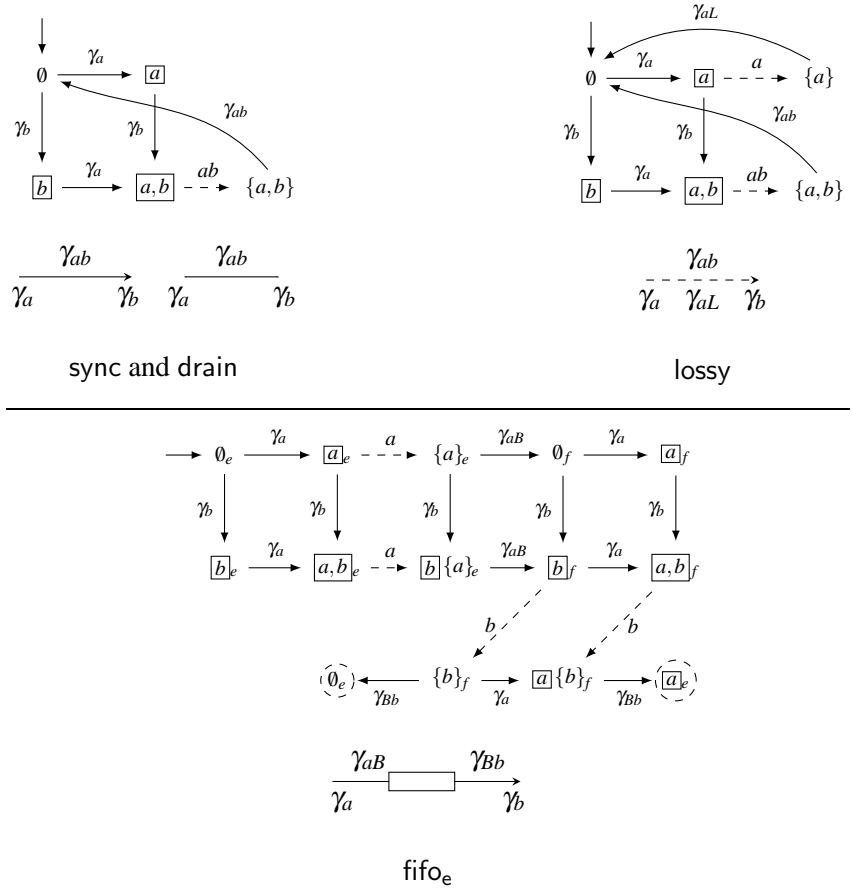

Figure~\ref{fig:imcreochannels} depicts the \imcreo models  corresponding to the basic \sreo channels. 
To simplify the picture, transition overlapping is generally avoided by the graphical replication of states suitably annotated with a dashed circle. 

The \imcreo model of a stochastic \reochannel{\syncname} channel is interpreted as follows: initially, no requests are pending neither in port $a$ nor in port $b$. Requests arrive at port $a$ (respectively, $b$) at rate  $\gamma_a$ (respectively, $\gamma_b$).
The channel \emph{blocks} until a request arrives to the other port. When state \enbox[\scriptsize]{a,b}  is reached, representing a configuration in which  both ports have pending requests, then both  eventually fire. 
That is, actions $a$ and $b$ are activated simultaneously. At this moment, the channel starts transmitting data between $a$ and $b$ and evolves back to the initial state with a processing delay rate of  
$\gamma_{ab}$.
For a stochastic \reochannel{\lossyname} channel the interpretation is similar. However it exhibits two additional transitions to model the possibility of data being lost: at state \enbox[\scriptsize]{a}, port $a$ may fire, because there is no pending request at port $b$. 
When such is the case, the channel evolves back to the initial state after a delay of discarding data.
State \enbox[\scriptsize]{a} captures the context-dependent behaviour characteristic of this channel.
Finally, the \reochannel{\fifoename} stochastic channel differs from the others by introducing an internal state. Notice how pending requests at port $a$ automatically fire when the \emph{buffer} is empty (states \enbox[\scriptsize]{a}$_e$ and \enbox[\scriptsize]{a,b}$_e$), and requests at port $b$ block until it is full (states \enbox[\scriptsize]{a,b}$_e$ and \enbox[\scriptsize]{b}\enset[\scriptsize]{a}$_e$).
Also, notice that, to maintain consistency, the internal state of this channel only changes after Markovian transitions,representing processing delays, succeed. Actually, this is the rule in \imcreo models.

The composition of  two  \imcreo models $I$ and $J$, with respect to a set of ports  $M \subseteq \nodeset$, is given by  a product (which accounts for parallel evolution) and a synchronisation operation (which deals with interaction), and denoted by
$$\partial_M (I_1\composevia I_2) $$
The definitions of both operations are collected in the appendix; the reader is referred to \cite{OSB15} for examples and details.
Note that this two-step composition approach is not a novelty in the definition of composition operations in \reo. Actually, it is  very much in the same spirit of the one defined for \reo automata~\cite{bonsangue2012}.

\section{\dimcreo: The new model}\label{sec:dimcreo}

As mentioned in the Introduction, \imcreo does not scale in a smooth way: composition  generates a state space that remains considerably big even after minimisation via bisimulation. This limits its use for analysis of coordination, namely in the context of cloud-based systems involving an arbitrary number of actors.

Actually, as an exogenous coordination model, \reo disregards services or components when it comes to specifying a coordination schema. It only assumes that such computation \emph{loci} are bound to the ports of the connector, which receive IO impulses whenever communication is requested. Consequently, \sreo inherits the same philosophy. But, does it? Not quite! 
In fact, \sreo circuits are not completely exogenous. They embody, in  request arrival rates, information that is inherently associated to the induced stochastic behaviour of the interacting services coordinated by \sreo circuits. 
As expected, this hampers the reutilisation of \sreo models, and introduces unnatural simplifications to make it compositional. 

As an alternative, we propose to consider the stochastic version of \reo as a two-phase component-based coordination model. The qualifier  \emph{two-phase}  stresses  the need for explicitly considering the model  before  and after deployment, known as the \emph{design} and \emph{deployment} phases, respectively; it is component-based because it is constructed from four specific components: the writer, the reader, the channel and the node, as graphically presented in Figure~\ref{fig:stochreo2_components}.

\begin{figure}[htbp]
\begin{center}

\tikz[baseline=-4]{
	\node[draw,rectangle,minimum size=0.15cm,fill=black] (a) at (0,0) {};
	\node[yshift=10pt] (rate) at (0,0)  {\footnotesize $\gamma_{wr}$};
}
\hspace{0.4cm}
\tikz[baseline=-4]{
	\node[draw,rectangle,minimum size=0.15cm] (a) at (0,0) {};
	\node[yshift=10pt] (rate) at (0,0)  {\footnotesize $\gamma_{rd}$};
}
\hspace{0.4cm}
\tikz{\sync{(0,0)}{(1,0)}{
		node[yshift=5pt] {\footnotesize $\gamma_{ab}$} 
		node[yshift=-6pt,xshift=-15pt] {\scriptsize $a$} 
		node[yshift=-5pt,xshift=15pt] {\scriptsize $b$} 
	}
}
\hspace{0.4cm}
\tikz[baseline=-4]{\mixednode{(m)}{(0,0)}{
		node[xshift=-10pt] {\footnotesize $\gamma_{e}$}
		node[xshift=10pt] {\footnotesize $\gamma_{d}$}	
	}
}

\caption[The essential components of \sreo]{The essential components of \sreo.}
\label{fig:stochreo2_components}
\end{center}
\end{figure}

The first two are synchronous stochastic abstractions of the real-world services that are to be bound to the ports of the connector. They are annotated with a delay rate ($\gamma_{wr}$ and $\gamma_{rd}$, respectively), that models the time between consecutive IO requests issued by them. 
The channel component inherits the usual behaviour of \reo channels, as well as  the  processing delay rate of \sreo, which models the duration of point-to-point data transportation. Note that the request arrival rates are no more part of a channel model.
The node is now taken as a synchronous component which behaves like the \reochannel{\replicatornodename}, the \reochannel{\mergernodename} or the \reochannel{\routernodename} connector.
Differently from the original version of \sreo, in this approach nodes are assumed to take time to enqueue and dequeue data. This behaviour is modelled by the delay rates $\gamma_e$ and $\gamma_d$:
\begin{itemize}
\item Enqueueing data takes into account not only the time to process incoming data but also the time needed to select from which channel data will be read (if a \reochannel\mergernodename);
\item  Dequeuing data takes into account the time to write data in the channels; it further comprises the time to generate copies of the data to write (in  a \reochannel\replicatornodename), and the time to decide to which channels it will write (in a \reochannel\routernodename). 
\end{itemize}
This captures a more realistic stochastic behaviour of nodes, as opposed to the usual \emph{self-contained pumping station} behavioural assumption.

\begin{figure}[htbp]
\begin{center}

\begin{minipage}[c]{0.45\linewidth}
\centering
\bf Design-phase model
\end{minipage}
\begin{minipage}[c]{0.45\linewidth}
\centering
\bf Deployment-phase model
\end{minipage}

\vspace{0.4cm}

\begin{minipage}[c]{0.45\linewidth}
\centering
\begin{tikzpicture}
	\mixednode{(m)}{(2,0)}{
		node[xshift=-5pt,yshift=8pt] {\tiny $\gamma_{e}$}
		node[xshift=5pt,yshift=-8pt] {\tiny $\gamma_{d}$}
	}
	\lossysync{(0,0)}{(m)}{
		node[yshift=8pt] {\footnotesize $\gamma_{ab}$}
		node[yshift=-8pt] {\footnotesize $\gamma_{aL}$}
		node[yshift=-8pt,xshift=-25pt] {\scriptsize $a$}
	}
	\fifoe{(m)}{(4,0)}{
		node[yshift=8pt,xshift=-30pt] {\footnotesize $\gamma_{bB}$}
		node[yshift=8pt,xshift=0pt] {\footnotesize $\gamma_{Bc}$}
		node[yshift=-8pt,xshift=10pt] {\scriptsize $c$}
	}
\end{tikzpicture}

\end{minipage}
\begin{minipage}[c]{0.45\linewidth}
\centering

\begin{tikzpicture}
	\reader{(r)}{(4,0)}{
		node[yshift=12pt,xshift=8pt]  {\footnotesize $\gamma_{rd_c}$}
	}
	\writer{(w)}{(0,0)}{
		node[yshift=12pt,xshift=-18pt]  {\footnotesize $\gamma_{wr_a}$}
	}
	\mixednode{(m)}{(2,0)}{
		node[xshift=-5pt,yshift=8pt] {\tiny $\gamma_{e}$}
		node[xshift=5pt,yshift=-8pt] {\tiny $\gamma_{d}$}
	}
	\lossysync{(w)}{(m)}{
		node[yshift=8pt] {\footnotesize $\gamma_{ab}$}
		node[yshift=-8pt] {\footnotesize $\gamma_{aL}$}
	}
	\fifoe{(m)}{(r)}{
		node[yshift=8pt,xshift=-30pt] {\footnotesize $\gamma_{bB}$}
		node[yshift=8pt,xshift=0pt] {\footnotesize $\gamma_{Bc}$}
	}
\end{tikzpicture}

\end{minipage}

\caption{The two-pase, component-based  model of a \reochannel{lossyfifo}.}
\label{fig:stochreo2}
\end{center}
\end{figure}
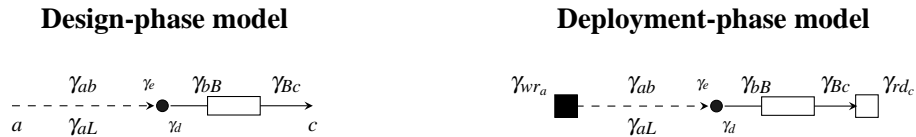

The design-phase  models come from the composition of channel and node components. In turn, deployment-phase models are fixed for a given installation of composed services. The writer and the reader components are bound to the interface ports of the connector. 
This is, in fact, very close to the original \sreo model, adding to it, however, a more realistic separation of concerns. 
Figure~\ref{fig:stochreo2} depicts a simple example of a \reochannel{lossyfifo} connector in both the design- and the deployment-phase.

This component-based rephrasing of \imcreo takes  each channel, node, writer and reader  as an independent stochastic process that may (or may not) synchronise with the other elements. 
The introduction of delays in nodes raises the need for two new sorts of states with specific semantics: the state where the node is enqueueing and the state where it is dequeueing data. A state in \dimcreo is fully characterised as $(R,T,E,D,Q)$ with $E,D \in \partsof{\nodeset}$, where states of the form $(\emptyset, \emptyset, E, \emptyset, Q)$ are \emph{enqueueing} states, \ie\ in which the node is reading from the channel ends in set E; these states are represented as $\enqueue[\scriptsize]{E}_Q$. Likewise, states of the form $(\emptyset, \emptyset, \emptyset, D, Q)$ are \emph{dequeueing} states, meaning that the node is writing to the channel ends in set $D$. These states are represented as $\dequeue[\scriptsize]{D}_Q$. 

Apart from this modification on states, the basic formal model of \imcreo  remains unchanged, as well as the variants of  bisimulation introduced in \cite{OSB15}. Let us, however, revisit the  \dimcreo for each basic component.

\paragraph{Channels.}
The \dimcreo models for the basic \reo channels are depicted in Figure~\ref{fig:imcreo_distilled_channels}. They are obtained from their counterpart in
 \imcreo models by disregarding the environment information. When compared to the corresponding \imcreo representation, a significant reduction is visible in their state space.

\begin{figure}[!htbp]
\begin{center}

\begin{minipage}[b]{0.32\linewidth}
\begin{center}
\tikz[>=latex, shorten >=1pt, node distance=1.5cm, style={font=\scriptsize}, initial where={above}, initial text=]{%
\node[initial] (e) {$\emptyset$}; %
\node (ab) [right of = e] {\entrans{a,b}} ; %
\path[->,dashed] (e) edge[bend left] node[above]{$ab$} (ab.140); 
\path[->] (ab) edge[bend left] node[below] {$\gamma_{ab}$} (e.290); 
}%
\\[0.3cm]
\tikz \sync{(0,0)}{(1.5,0)}{
	node[above]{$\gamma_{ab}$} 
	node[below,xshift=-20pt]{\scriptsize$a$}
	node[below,xshift=20pt]{\scriptsize$b$} 
};
\tikz \syncdrain{(0,0)}{(1.5,0)}{
	node[above]{$\gamma_{ab}$} 
	node[below,xshift=-20pt]{\scriptsize$a$} 
	node[below,xshift=20pt]{\scriptsize$b$}
};
\\[0.3cm]
\reochannel{\syncname} and \reochannel{\drainname}
\end{center}
\end{minipage}
\begin{minipage}[b]{0.32\linewidth}
\begin{center}
\tikz[>=latex, shorten >=1pt, node distance=1.5cm, style={font=\scriptsize}, initial where={above}, initial text=]{%
\node[initial] (e) {$\emptyset$}; %
\node (a) at ($(e) + (2,.8) $) {\entrans{a}} ; %
\node (ab) at ($(e) + (2,-.8) $) {\entrans{a,b}} ; %
\path[->,dashed] (e) edge node[above]{$a$} (a); 
\path[->,dashed] (e) edge node[above]{$ab$} (ab); 
\path[->] (a) edge[bend right] node[above] {$\gamma_{aL}$} (e.70);
\path[->] (ab) edge[bend left] node[below] {$\gamma_{ab}$} (e.290); 
}%
\\[0.3cm]
\tikz \lossysync{(0,0)}{(1.5,0)}{
	node[above]{$\gamma_{ab}$} 
	node[below]{$\gamma_{aL}$} 
	node[below,xshift=-20pt]{\scriptsize$a$} 
	node[below,xshift=20pt]{\scriptsize$b$}
};\\[0.3cm]
\reochannel{\lossyname}
\end{center}
\end{minipage}
\begin{minipage}[b]{0.32\linewidth}
\begin{center}
\tikz[>=latex, shorten >=1pt, node distance=1.5cm, style={font=\scriptsize}, initial where={above}, initial text=]{%
\node[initial] (e) {$\emptyset_e$}; %
\node (a) [right of = e] {$\entrans{a}_e$} ; %
\node (ef) [below of = a] {$\emptyset_f$} ; %
\node (bf) [left of = ef] {$\entrans{b}_f$} ; %
\path[->,dashed] (e) edge node[above]{$a$} (a); %
\path[->] (a) edge node[right]{$\gamma_{aB}$} (ef); %
\path[->,dashed] (ef) edge node[below] {$b$} (bf); %
\path[->] (bf) edge node[left] {$\gamma_{Bb}$} (e); %
}%
\\[0.3cm]
\tikz \fifoe{(0,0)}{(2,0)}{
	node[above,xshift=-35pt]{$\gamma_{aB}$} 
	node[above]{$\gamma_{Bb}$} 
	node[below,xshift=-45pt]{\scriptsize$a$} 
	node[below,xshift=10pt]{\scriptsize$b$}
};\\[0.3cm]\reochannel{\fifoename}
\end{center}
\end{minipage}

\caption{The \dimcreo models for  basic \sreo channels.}
\label{fig:imcreo_distilled_channels}
\end{center}
\end{figure}

 \begin{figure}[!htbp]
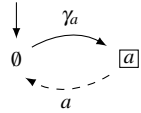

\begin{center}
\tikz[>=latex, shorten >=1pt, node distance=1.5cm, style={font=\scriptsize}, initial where={above}, initial text=]{%
\node[initial] (e) {$\emptyset$}; %
\node (a) [right of = e] {\enbox{a}} ; %
\path[->] (e) edge[bend left] node[above]{$\gamma_{a}$} (a.150); 
\path[->,dashed] (a) edge[bend left] node[below] {$a$} (e.290); 
}%
\caption{The \dimcreo for the reader and writer components}
\label{fig:imcreo_distilled_readerwriter}
\end{center}
\end{figure}

\paragraph{Readers and writers.}
To obtain deployment-phase models it is necessary to compose design-phase models with the environment information, \ie\ the reader and the writer components. Observationally, the latter would behave similarly: they issue IO requests by publishing the intention to write (respectively, read) data; then they  block until synchronising with the connector ports. Thus, one single \dimcreo model is enough to capture such behaviour, as depicted in 
Figure~\ref{fig:imcreo_distilled_readerwriter}.

A reader is bound to an output port while a writer is bound to an input port. This is how readers and writers are distinguished.
The composition of these components with one channel will result in a \dimcreo model capturing the semantics of \sreo channels (and consequently, connectors). 

\begin{figure}[htbp]
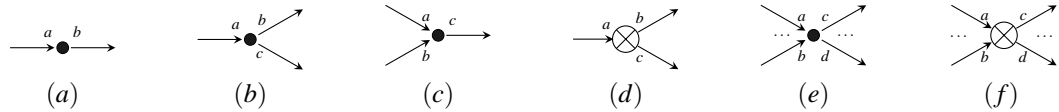

\begin{center}
\begin{minipage}[c]{0.15\linewidth}
\centering
\tikz{
	\mixednode{(m)}{(0.7,0)}{}
	\sync{(0,0)}{(m)}{
		node[above,xshift=6pt]{\tiny $a$}
	}
	\sync{(m)}{(1.4,0)}{
		node[above,xshift=-6pt]{\tiny $b$}
	}
}
\end{minipage}
\begin{minipage}[c]{0.15\linewidth}
\centering
\tikz{
	\mixednode{(m)}{(0.7,0)}{}
	\sync{(0,0)}{(m)}{
		node[above,xshift=6pt]{\tiny $a$}
	}
	\sync{(m)}{(1.4,0.4)}{
		node[above,xshift=-7pt,yshift=-5pt]{\tiny $b$}
	}
	\sync{(m)}{(1.4,-0.4)}{
		node[below,xshift=-8pt,yshift=6pt]{\tiny $c$}
	}
}
\end{minipage}
\begin{minipage}[c]{0.15\linewidth}
\centering
\tikz{
	\mixednode{(m)}{(0.7,0)}{}
	\sync{(0,0.4)}{(m)}{
		node[above,xshift=7pt,yshift=-5pt]{\tiny $a$}
	}
	\sync{(0,-0.4)}{(m)}{
		node[below,xshift=7pt,yshift=5pt]{\tiny $b$}
	}
	\sync{(m)}{(1.4,0)}{
		node[above,xshift=-6pt]{\tiny $c$}
	}
}
\end{minipage}
\begin{minipage}[c]{0.15\linewidth}
\centering
\tikz{
	\xrouter{(m)}{(0.7,0)}{}
	\sync{(0,0)}{(m)}{
		node[above,xshift=5pt]{\tiny $a$}
	}
	\sync{(m)}{(1.4,0.4)}{
		node[above,xshift=-7pt,yshift=-5pt]{\tiny $b$}
	}
	\sync{(m)}{(1.4,-0.4)}{
		node[below,xshift=-7pt,yshift=5pt]{\tiny $c$}
	}
}
\end{minipage}
\begin{minipage}[c]{0.15\linewidth}
\centering
\tikz{
	\mixednode{(m)}{(0.7,0)}{
		node[left, xshift=-5pt] {\tiny $\ldots$}
		node[right, xshift=5pt] {\tiny $\ldots$}
	}
	\sync{(0,0.4)}{(m)}{
		node[above,xshift=7pt,yshift=-5pt]{\tiny $a$}
	}
	\sync{(0,-0.4)}{(m)}{
		node[below,xshift=7pt,yshift=5pt]{\tiny $b$}
	}
	\sync{(m)}{(1.4,0.4)}{
		node[above,xshift=-7pt,yshift=-5pt]{\tiny $c$}
	}
	\sync{(m)}{(1.4,-0.4)}{
		node[below,xshift=-7pt,yshift=5pt]{\tiny $d$}
	}
}
\end{minipage}
\begin{minipage}[c]{0.15\linewidth}
\centering
\tikz{
	\xrouter{(m)}{(0.7,0)}{}
	\sync{(0,0.4)}{(m)}{
		node[below, yshift=-3pt,xshift=-5pt] {\tiny $\ldots$}
		node[above,xshift=5pt,yshift=-5pt]{\tiny $a$}
	}
	\sync{(0,-0.4)}{(m)}{
		node[below,xshift=5pt,yshift=5pt]{\tiny $b$}
	}
	\sync{(m)}{(1.4,0.4)}{
		node[below, yshift=-3pt,xshift=3pt] {\tiny $\ldots$}
		node[above,xshift=-5pt,yshift=-5pt]{\tiny $c$}
	}
	\sync{(m)}{(1.4,-0.4)}{
		node[below,xshift=-5pt,yshift=5pt]{\tiny $d$}
	}
}
\end{minipage}

\vspace{0.1cm}

\begin{minipage}[c]{0.15\linewidth}
\centering
\small $(a)$
\end{minipage}
\begin{minipage}[c]{0.15\linewidth}
\centering
\small $(b)$
\end{minipage}
\begin{minipage}[c]{0.15\linewidth}
\centering
\small $(c)$
\end{minipage}
\begin{minipage}[c]{0.15\linewidth}
\centering
\small $(d)$
\end{minipage}
\begin{minipage}[c]{0.15\linewidth}
\centering
\small $(e)$
\end{minipage}
\begin{minipage}[c]{0.15\linewidth}
\centering
\small $(f)$
\end{minipage}

\caption[Different \reo node configurations.]{$(a)$ simple; $(b)$ replicator; $(c)$ merger; $(d)$ router$; (e)$ merger-replicator: $(f)$ merger-router.}
\label{fig:node_configurations}
\end{center}
\end{figure}

\paragraph{Nodes.}
The basic \reo node ontology (\reochannel{\replicatornodename}, \reochannel{\mergernodename} and \reochannel{\routernodename})
is extended to the six different  configurations based over them, as shown in Figure~\ref{fig:node_configurations}.

Note that node configurations $(a)$ to $(c)$ are special cases of $(e)$: these nodes select one incoming channel to read data from, and then copy and write the data into all the outgoing channels. In turn, node configuration $(d)$ is a special case of $(f)$: it selects one incoming channel to read from, and then routes the data to one of the outgoing channels. Nodes $(e)$ and $(f)$ define, in fact, two families of nodes, referred henceforth as \reochannel\mrname and \reochannel\mxname, respectively. They are parametric on the number of incoming and outgoing channels and also on the delays for reading (enqueueing) and writing (dequeueing) data, whenever such delays are considered.
Consistently, all \dimcreo nodes are generated from these two families, taking into account their parameters as follows:  
$$
\reochannel\mrname, \reochannel\mxname :\,  \partsof{\nodeset} \times \partsof{\nodeset} \times \realsetpos \times \realsetpos 
$$
where the first parameter is a set of output channel ends (the node inputs); the second is a set of input channel ends (its outputs); the third models the time to select and read from one channel end, and finally, the fourth parameter models the time to copy, route and write data into one channel end.

Figure~\ref{fig:nodes_imcreo_model} depicts the parametric \dimcreo models for both the \reochannel{\mrname} and the \reochannel{\mxname} families of nodes.
 Notation $I_i$ represents the $i^{th}$ element in set $I$ and $\overline{O}$ represents the concatenation of all elements in set $O$. Moreover, it is assumed that the cardinality of sets $I$ and $O$ are, respectively, $n$ and $k$.

\begin{figure}[!htbp]
\begin{center}

\begin{minipage}[b]{0.48\linewidth}
\begin{center}

\tikz[>=latex, shorten >=1pt, node distance=1.5cm, style={font=\scriptsize}, initial where={left}, initial text=]{%
\node[initial] (e) {$\emptyset$}; %
\node (i0) at ($(e) + (2,.8)$) {\enqueue{I_1}} ; %
\node (ith1) at ($(e) + (1,0)$) {\ldots} ; %
\node (in) at ($(e) + (2,-.8)$) {\enqueue{I_{n}}} ; %
\node (ith2) at ($(e) + (2,0)$) {\ldots} ; %
\node (o) at ($(e) + (4,0)$) {\dequeue{O}} ; %
\node (ith3) at ($(e) + (3,0)$) {\ldots} ; %
\node (_efict) [right of = o] {$\emptyset$} ;
\encircle[-1mm]{efict}{_efict};
\path[->,dashed] (e) edge[out=0,in=180] node[above, xshift=-13pt]{$I_1\overline{O}$} (i0); 
\path[->,dashed] (e) edge[out=0,in=180] node[below, xshift=-19pt] {$I_{n}\overline{O}$} (in);
\path[->] (i0) edge[out=0,in=180] node[above] {$\gamma_{e}$} (o);
\path[->] (in) edge[out=0,in=180] node[below] {$\gamma_{e}$} (o);
\path[->] (o) edge node[above] {$\frac{\gamma_d}{k}$} (efict);
}%
\\[0.85cm]
\reochannel{\mrname}$(I,O,\gamma_e,\gamma_d)$

\end{center}
\end{minipage}
\begin{minipage}[b]{0.48\linewidth}
\begin{center}

\tikz[>=latex, shorten >=1pt, node distance=1.5cm, style={font=\scriptsize}, initial where={left}, initial text=]{%
\node[initial] (e) {$\emptyset$}; %
\node (i0o0) at ($(e) + (2,1.6)$) {\enqueue{I_1,O_1}} ; %
\node (kth0) at ($(e) + (2,1.2)$) {\ldots} ; %
\node (i0ok) at ($(e) + (2,.8)$) {\enqueue{I_1,O_k}} ; %
\node (kth1) at ($(e) + (.7,1)$) {\ldots} ; %
\node (kth2) at ($(e) + (.7,-1)$) {\ldots} ; %
\node (ith1) at ($(e) + (1,0)$) {\ldots} ; %
\node (ino0) at ($(e) + (2,-.8)$) {\enqueue{I_{n},O_1}} ; %
\node (ith0) at ($(e) + (2,-1.2)$) {\ldots} ; %
\node (inok) at ($(e) + (2,-1.6)$) {\enqueue{I_{n},O_k}} ; %
\node (ith2) at ($(e) + (2,0)$) {\ldots} ; %
\node (o0) at ($(e) + (4,.8)$) {\dequeue{O_1}} ; %
\node (ok) at ($(e) + (4,-.8)$) {\dequeue{O_{k}}} ; %
\node (kth3) at ($(e) + (4,0)$) {\ldots} ; %
\node (ith3) at ($(e) + (3,0)$) {\ldots} ; %
\node (_efict) at ($(e) + (5.5,0)$) {$\emptyset$} ;
\encircle[-1mm]{efict}{_efict};
\path[->,dashed] (e) edge[out=90,in=180] node[above, xshift=-10pt]{$I_1O_1$} (i0o0); 
\path[->,dashed] (e) edge[out=0,in=180] node[below, yshift=5pt,xshift=12pt]{$I_1O_{k}$} (i0ok); 
\path[->,dashed] (e) edge[out=0,in=180] node[above, yshift=-5pt,xshift=12pt] {$I_{n}O_{1}$} (ino0);
\path[->,dashed] (e) edge[out=270,in=180] node[below, xshift=-10pt] {$I_{n}O_{k}$} (inok);
\path[->] (i0o0) edge[out=0,in=90] node[above] {$\gamma_{e}$} (o0);
\path[->] (i0ok) edge[out=0,in=110] node[below, xshift=-3,yshift=3] {$\gamma_{e}$} (ok);
\path[->] (ino0) edge[out=0,in=250] node[above,xshift=-3,yshift=-3] {$\gamma_{e}$} (o0);
\path[->] (inok) edge[out=0,in=270] node[below] {$\gamma_{e}$} (ok);
\path[->] (o0) edge node[above] {$\gamma_{d}$} (efict);
\path[->] (ok) edge node[above] {$\gamma_{d}$} (efict);
}%
\\[0.1cm]
 \reochannel{\mxname}$(I,O,\gamma_e,\gamma_d)$

\end{center}
\end{minipage}

\caption{\dimcreo models for \reochannel{\mrname} and \reochannel{\mxname} nodes.}
\label{fig:nodes_imcreo_model}
\end{center}
\end{figure}

The \reochannel{\mrname} node blocks until synchronising with one of the input channel ends and all the output channel ends. On synchronisation, it starts enqueueing data from the input channel end (delayed for some exponentially distributed time modelled by $\gamma_e$). Then, it dequeues data to all the output channel ends and returns to the initial blocked state. The delay time of a single dequeue operation is exponentially distributed with rate $\gamma_d$; since it performs $k$ such operations, then the average delaying time is exponentially distributed with rate $\frac{\gamma_d}{k}$.
The \reochannel{\mxname}, in turn, blocks until synchronising with one of the input and one of the output channels ends. On synchronisation, it goes to an enqueueing state and remains there for an exponentially distributed time modelled by rate $\gamma_e$. Then, it dequeues data to the selected output channel end at a rate $\gamma_d$, returning to the initial blocked state.

By disregarding enqueueing and dequeueing delays, these families of nodes are simplified into a single \dimcreo model with transition space size of $n$ and $n.k$ for \reochannel\mrname and \reochannel\mxname, respectively, corresponding only to the interactive transitions. 

\section{Composition in \dimcreo}\label{sec:composition_dimcreo}

Composition in \dimcreo extends that of  \imcreo, adding to the parallel and synchronization steps (see Appendix),  a phase for \emph{cleaning} superfluous transitions which takes into account the need for  enqueueing/dequeueing data in a specific order. Concretely, $(i)$ data is always enqueued into the node only after being transmitted to that node; $(ii)$ data is always transmitted to any further node only after being dequeued from the current one and $(iii)$ data is always enqueued before being dequeued (from the same node). Actually, \dimcreo requires that enqueueing and dequeueing transitions appear immediately one after the other, except in cases where other operations may occur in parallel; when such is the case, transitions will appear interleaved.
Formally, the  cleaning operation is defined as follows:

\begin{definition}[\dimcreo clean up]
Let $M \subseteq \nodeset$ and $I=(S, Act, \intertrans, \marktrans, s)$ be a \dimcreo.  Assume also  a relation $<$ on $\nodeset$ such that $a < b$ when data flows from $a$ to $b$, with  $a,b\in \nodeset$, which is lifted to sets as expected:  $A < B \sse \exists_{a\in A} \ \cdot\ \forall_{b\in B} \ \cdot \ a < b$.

The cleaning of $I$ with respect to $M$, denoted $\mathcal{C}_M I$, corresponds to restricting $\partial_M I$ so that all its Markovian transitions $i \stackrel \gamma \marktrans f$ respect:
\begin{itemize}
\item[(i)] $R_f \cap AN(i) = \emptyset $, where $AN(i) = T_i \cup \setdef{j \in \nodeset}{\exists_{k \in T_i} . j <k \lor k<j}$;
\item[(ii)] $
\left\{
\begin{array}{l l}
    T_i = T_f 					& \quad \text{if} \quad E_i < T_i  \quad \text{or} \quad  T_i \cap D_i \neq \emptyset \\
    T_i\setminus T_f < T_f		& \quad \text{otherwise}
\end{array} \right.
$
\end{itemize}
and all its interactive transitions $j \stackrel X \intertrans k$ respect:
\begin{itemize}
\item[(iii)] $\neg \exists_{j \stackrel Y \intertrans l \in \intertrans} \ \cdot \ X=Y \land T_k \cap M = \emptyset \land T_l \cap M \neq \emptyset$.
\end{itemize}

\end{definition}

\begin{figure}[!htbp]
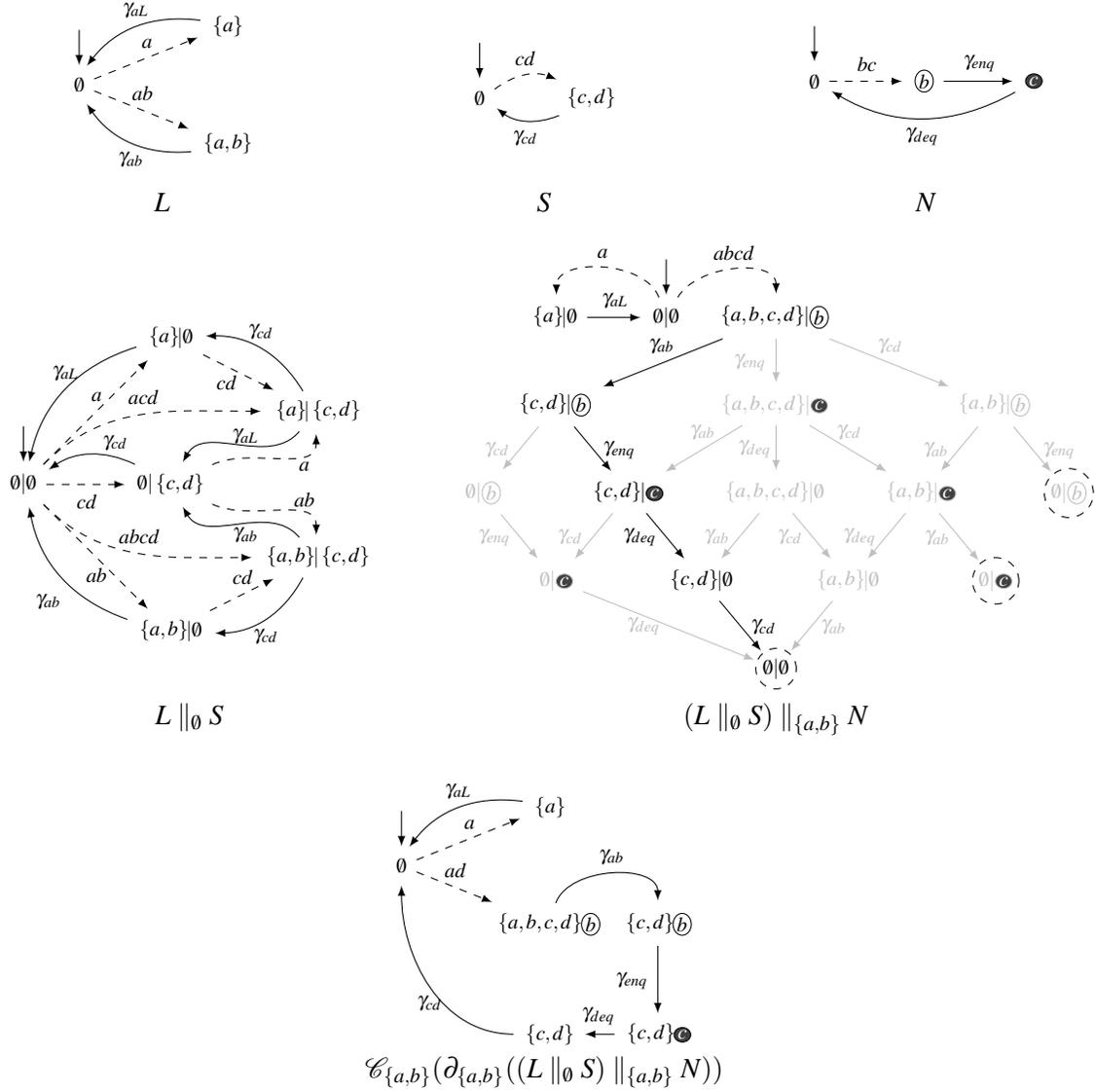

\begin{center}
\begin{minipage}[b]{0.32\linewidth}
\begin{center}
\tikz[>=latex, shorten >=1pt, node distance=1.5cm, style={font=\scriptsize}, initial where={above}, initial text=]{%
\node[initial] (e) {$\emptyset$}; %
\node (a) at ($(e) + (2,.8) $) {\entrans{a}} ; %
\node (ab) at ($(e) + (2,-.8) $) {\entrans{a,b}} ; %
\path[->,dashed] (e) edge node[above]{$a$} (a); 
\path[->,dashed] (e) edge node[above]{$ab$} (ab); 
\path[->] (a) edge[bend right] node[above] {$\gamma_{aL}$} (e.70);
\path[->] (ab) edge[bend left] node[below] {$\gamma_{ab}$} (e.290); 
}\\
$L$
\end{center}
\end{minipage}
\begin{minipage}[b]{0.32\linewidth}
\begin{center}

\tikz[>=latex, shorten >=1pt, node distance=1.5cm, style={font=\scriptsize}, initial where={above}, initial text=]{%
\node[initial] (e) {$\emptyset$}; %
\node (cd) [right of = e] {\entrans{c,d}} ;
\path[->,dashed] (e) edge[bend left] node[above]{$cd$} (cd)  ;
\path[->] (cd) edge[bend left] node[below]{$\gamma_{cd}$} (e)  ;
}\\[0.3cm]
$S$
\end{center}
\end{minipage}
\begin{minipage}[b]{0.32\linewidth}
\begin{center}

\tikz[>=latex, shorten >=1pt, node distance=1.5cm, style={font=\scriptsize}, initial where={above}, initial text=]{%
\node[initial] (e) {$\emptyset$}; %
\node (benq) [right of = e] {\enqueue{b}} ;
\node (cdeq) [right of = benq] {\dequeue{c}} ;
\path[->,dashed] (e) edge node[above]{$bc$} (benq)  ;
\path[->] (benq) edge node[above]{$\gamma_{enq}$} (cdeq)  ;
\path[->] (cdeq) edge[bend left] node[below]{$\gamma_{deq}$} (e)  ;
}\\[0.3cm]
$N$
\end{center}
\end{minipage}

\vspace{0.3cm}

\begin{minipage}[b]{0.39\linewidth}
\begin{center}
\tikz[>=latex, shorten >=1pt, node distance=1.5cm, style={font=\scriptsize}, initial where={above}, initial text=]{%
\node[initial] (e) {$\emptyset|\emptyset$}; %
\node (atrs) at ($(e) + (2,2)$) {\entrans{a}$|\emptyset$} ;
\node (cdtrs) at ($(e) + (2,0)$) {$\emptyset|$\entrans{c,d}} ;
\node (abtrs) at ($(e) + (2,-2)$) {\entrans{a,b}$|\emptyset$} ;
\node (acdtrs) at ($(e) + (4,1)$) {\entrans{a}$|$\entrans{c,d}} ;
\node (abcdtrs) at ($(e) + (4,-1)$) {\entrans{a,b}$|$\entrans{c,d}} ;
\path[->,dashed] (e) edge node[above]{$a$} (atrs);
\path[->,dashed] (e) edge[out=45,in=180] node[above]{$acd$} (acdtrs);
\path[->,dashed] (e) edge node[below]{$cd$} (cdtrs);
\path[->,dashed] (e) edge[out=-45,in=180] node[above]{$abcd$} (abcdtrs);
\path[->,dashed] (e) edge node[below]{$ab$} (abtrs);
\path[->,dashed] (atrs) edge node[below,xshift=-5pt,yshift=3pt]{$cd$} (acdtrs);
\path[->] (atrs) edge[bend right] node[above]{$\gamma_{aL}$} (e);
\path[->,dashed] (cdtrs) edge[out=25,in=270] node[below,xshift=10pt,yshift=2pt]{$a$} (acdtrs);
\path[->,dashed] (cdtrs) edge[out=-25,in=90] node[above,xshift=10pt,yshift=-2pt]{$ab$} (abcdtrs);
\path[->] (cdtrs) edge[bend right] node[above,xshift=10pt,yshift=-2pt]{$\gamma_{cd}$} (e);
\path[->,dashed] (abtrs) edge node[above]{$cd$} (abcdtrs);
\path[->] (abtrs) edge[bend left] node[below,xshift=-5pt,yshift=-3pt]{$\gamma_{ab}$} (e);
\path[->] (abcdtrs) edge[bend left] node[below]{$\gamma_{cd}$} (abtrs);
\path[->] (abcdtrs) edge[out=130,in=-60] node[below,xshift=3pt,yshift=2pt]{$\gamma_{ab}$} (cdtrs);
\path[->] (acdtrs) edge[out=-130,in=60] node[above,xshift=3pt,yshift=-3pt]{$\gamma_{aL}$} (cdtrs);
\path[->] (acdtrs) edge[bend right] node[above]{$\gamma_{cd}$} (atrs);
}\\[0.5cm]
$L\parallel_{\emptyset}S$
\end{center}
\end{minipage}
\begin{minipage}[b]{0.6\linewidth}
\begin{center}

\tikz[>=latex, shorten >=1pt, node distance=1.5cm, style={font=\scriptsize}, initial where={above}, initial text=]{%
\node[initial] (e) {$\emptyset|\emptyset$} ;
\node (a) [left of = e] {$\entrans{a}|\emptyset$} ;
\node (abcdbenq) [right of = e] {$\entrans{a,b,c,d}|\enqueue{b}$} ;
\node (cdbenq) at ($(abcdbenq) + (-3,-1.2)$) {$\entrans{c,d}|\enqueue{b}$} ;
\node (abcdcdeq) at ($(abcdbenq) + (0,-1.2)$) {\color{gray!50!white}$\entrans{a,b,c,d}|\dequeue{c}$} ;
\node (abbenq) at ($(abcdbenq) + (3,-1.2)$) {\color{gray!50!white}$\entrans{a,b}|\enqueue{b}$} ;
\node (benq) at ($(cdbenq) + (-1,-1.2)$) {\color{gray!50!white}$\emptyset|\enqueue{b}$} ;
\node (cdcdeq) at ($(cdbenq) + (1,-1.2)$) {$\entrans{c,d}|\dequeue{c}$} ;
\node (abcd) at ($(abcdcdeq) + (0,-1.2)$) {\color{gray!50!white}$\entrans{a,b,c,d}|\emptyset$} ;
\node (abcdeq) at ($(abbenq) + (-1,-1.2)$) {\color{gray!50!white}$\entrans{a,b}|\dequeue{c}$} ;
\node (_benq) at ($(abbenq) + (1,-1.2)$) {\color{gray!50!white}$\emptyset|\enqueue{b}$} ;
\encircle[-1mm]{benqfict}{_benq} ;
\node (cdeq) at ($(cdbenq) + (0,-2.4)$) {\color{gray!50!white}$\emptyset|\dequeue{c}$} ;
\node (cd) at ($(cdcdeq) + (1,-1.2)$) {$\entrans{c,d}|\emptyset$} ;
\node (ab) at ($(abcd) + (1,-1.2)$) {\color{gray!50!white}$\entrans{a,b}|\emptyset$} ;
\node (_cdeq) at ($(abbenq) + (0,-2.4)$) {\color{gray!50!white}$\emptyset|\dequeue{c}$} ;
\encircle[-1mm]{cdeqfict}{_cdeq}
\node (_e) at ($(abcd) + (0,-2.4)$) {$\emptyset|\emptyset$} ;
\encircle[-1mm]{efict}{_e}
%
\path[->,dashed] (e) edge[out=110,in=90] node[above]{$a$} (a);
\path[->,dashed] (e) edge[out=60,in=90] node[above]{$abcd$} (abcdbenq); 
\path[->] (a) edge node[above]{$\gamma_{aL}$} (e) ;
\path[->] (abcdbenq) edge node[above]{$\gamma_{ab}$} (cdbenq);
\path[->,gray!50!white] (abcdbenq) edge node[left]{$\gamma_{enq}$} (abcdcdeq);
\path[->,gray!50!white] (abcdbenq) edge node[above]{$\gamma_{cd}$} (abbenq);
\path[->,gray!50!white] (cdbenq) edge node[left]{$\gamma_{cd}$} (benq);
\path[->] (cdbenq) edge node[right]{$\gamma_{enq}$} (cdcdeq);
\path[->,gray!50!white] (abcdcdeq) edge node[above]{$\gamma_{ab}$} (cdcdeq);
\path[->,gray!50!white] (abcdcdeq) edge node[left,xshift=2pt]{$\gamma_{deq}$} (abcd);
\path[->,gray!50!white] (abcdcdeq) edge node[above]{$\gamma_{cd}$} (abcdeq);
\path[->,gray!50!white] (abbenq) edge node[left]{$\gamma_{ab}$} (abcdeq);
\path[->,gray!50!white] (abbenq) edge node[right]{$\gamma_{enq}$} (benqfict);
\path[->,gray!50!white] (benq) edge node[left]{$\gamma_{enq}$} (cdeq);
\path[->,gray!50!white] (cdcdeq) edge node[left]{$\gamma_{cd}$} (cdeq);
\path[->] (cdcdeq) edge node[left]{$\gamma_{deq}$} (cd);
\path[->,gray!50!white] (abcd) edge node[left]{$\gamma_{ab}$} (cd);
\path[->,gray!50!white] (abcd) edge node[left]{$\gamma_{cd}$} (ab);
\path[->,gray!50!white] (abcdeq) edge node[left]{$\gamma_{deq}$} (ab);
\path[->,gray!50!white] (abcdeq) edge node[left]{$\gamma_{ab}$} (cdeqfict);
\path[->,gray!50!white] (cdeq) edge node[left]{$\gamma_{deq}$} (efict);
\path[->] (cd) edge node[right]{$\gamma_{cd}$} (efict);
\path[->,gray!50!white] (ab) edge node[right]{$\gamma_{ab}$} (efict);
}\\
$(L\parallel_{\emptyset}S)\parallel_{\enset{a,b}}N$
\end{center}
\end{minipage}

\vspace{0.5cm}

\begin{minipage}{\linewidth}
\centering
\tikz[>=latex, shorten >=1pt, node distance=1.5cm, style={font=\scriptsize}, initial where={above}, initial text=]{%
\node[initial] (e) {$\emptyset$}; %
\node (a) at ($(e) + (2,.8) $) {\entrans{a}} ; %
\node (abcdbenq) at ($(e) + (2,-.8) $) {\entrans{a,b,c,d}\enqueue{b}} ; %
\node (cdbenq) [right of = abcdbenq] {\entrans{c,d}\enqueue{b}} ; %
\node (cdcdeq) [below of = cdbenq] {\entrans{c,d}\dequeue{c}} ; %
\node (cdeq) [left of = cdcdeq] {\entrans{c,d}} ; %
\path[->,dashed] (e) edge node[above]{$a$} (a); 
\path[->,dashed] (e) edge node[above]{$ad$} (abcdbenq); 
\path[->] (a) edge[bend right] node[above] {$\gamma_{aL}$} (e.70);
\path[->] (abcdbenq) edge[out=70,in=90] node[above] {$\gamma_{ab}$} (cdbenq); 
\path[->] (cdbenq) edge node[left] {$\gamma_{enq}$} (cdcdeq); 
\path[->] (cdcdeq) edge node[above] {$\gamma_{deq}$} (cdeq); 
\path[->] (cdeq) edge[out=180,in=270] node[below] {$\gamma_{cd}$} (e); 
}\\[-1cm]
$$\mathcal{C}_{\enset{a,b}}(\partial_{\enset{a,b}}((L\parallel_{\emptyset}S)\parallel_{\enset{a,b}}N ) ) $$
\end{minipage}

\caption[Design-phase \dimcreo model for the \reochannel{lossysync} connector.]{Design-phase of a \dimcreo model for the \reochannel{lossysync} connector with a delayed node.}
\label{fig:distilled_lossysync}
\end{center}
\end{figure}

The following example shows the (design-phase) composition of a \reochannel{\lossyname} channel with a \reochannel{\syncname} channel, considering that data enqueueing and dequeueing  in the mixed node is delayed with rates $\gamma_{enq}$ and $\gamma_{deq}$, respectively. Figure~\ref{fig:distilled_lossysync} depicts the composition of the two channels and the synchronising node. The greyed-out transitions are eliminated by  cleaning, as they fail to respect sequencing.

\medskip

In order to obtain the deployment-phase model, an extra step is required that composes the design-phase model with the environment model. Formally,

\begin{definition}[Deployment]
Let $I$ be a \dimcreo model of a design-phase connector, $E$ a set of \dimcreo models representing all the relevant reader and writer components defining the environment for I, and finally $M\subseteq\nodeset$.
The deployment-phase model of $I$ in environment $E$ with respect to the set of ports $M$ is computed by 
$$
\mathcal{C}_M (I \parallel_M E_{\parallel}),
$$
where $E_\parallel$ is the parallel composition of all elements of $E$, referred to as the global environment model.
\end{definition}

Note that whenever nodes do not delay the system, composition of \dimcreo models are boiled down to that defined for \imcreo. This is stated formally 
 in the following theorem proved in \cite{Olitese13}:

\begin{theorem}\label{teo:imcreo_dimcreo_related}
Let $I$ be an \imcreo and $J$ a deployed \dimcreo. Consider that both $I$ and $J$ model the same \sreo connector (\ie\ with same stochastic information for channels and environment). Then, $I \bisim J \ \sse$ $J$ has no enqueuing and dequeueing states.
\end{theorem}

\section{Concluding}

This paper introduced \dimcreo --- a model for \sreo based on interactive Markov chains, which extends our previous work on \imcreo, increasing its scalability while retaining expressivity and compositionality. We believe coordination models are a major area of application of  formal models to cloud applications, with an enormous potential for their correct design and analysis.

This debate, however, is still in its infancy; only time and experience with real, challenging application, will provide sustainable evidence for the claim made here, as well as for the approach proposed.

\bibliographystyle{eptcs}


\section*{Appendix - Composition in \imcreo}

In \cite{OSB15} the composition of two \imcreo models $I_1$ and $I_2$, with respect to $M \subseteq \nodeset$, is given by   
$$\partial_M (I_1\composevia I_2) $$
comprising a \emph{product} and a \emph{synchronization} operator. This appendix recalls the corresponding definition.

\begin{definition}[Parallel Composition]\label{def:composition}
Let $I = (S_I, Act_I, \intertrans_I, \marktrans_I, s_i)$ and $J = (S_J, Act_J, \intertrans_J, \marktrans_J, s_j)$ be two \imcreo models. The
 parallel composition of $I$ and $J$ with respect to a set $M \subseteq \nodeset$ is defined as   
$$I \composevia J = (S, Act, \intertrans, \marktrans, (s_i, s_j))$$
where $S = S_I \times S_J$, $Act = Act_I \cup Act_J$, and \intertrans \ and \marktrans \ are the smallest relations satisfying

\begin{minipage}[m]{0.45\linewidth}
\small
$$
1. \quad \frac{i_1 \stackrel{A_I}{\intertrans_I} i_2  \quad A_I \cap M = \emptyset}%
		 {(i_1, j) \stackrel{A_I}{\intertrans} (i_2, j), \ \textnormal{for} \ j \in S_J}%
$$
\end{minipage}
\begin{minipage}[m]{0.45\linewidth}
\small
$$
2. \quad \frac{j_1 \stackrel{A_J}{\intertrans_J} j_2  \quad A_J \cap M = \emptyset }%
		 	  {(i, j_1) \stackrel{A_J}{\intertrans} (i, j_2), \ \textnormal{for} \ i \in S_I}%
$$
\end{minipage}

\vspace{0.6cm}

\begin{minipage}[m]{\linewidth}
\small
$$
3. \quad \frac{i_1 \stackrel{A_I}{\intertrans_I} i_2 \quad j_1 \stackrel{A_J}{\intertrans_J} j_2 \quad (A_I \cap A_J) \subseteq M \quad A_I,A_J \neq \emptyset}%
		 	  {(i_1, j_1) \stackrel{A_I \cup A_J}{\intertrans} (i_2, j_2)}%
$$
\end{minipage}

\begin{minipage}[m]{0.45\linewidth}
\small
$$
4. \quad \frac{i_1 \stackrel{\gamma}{\marktrans_I} i_2}%
		 	  {(i_1, j) \stackrel{\gamma}{\marktrans} (i_2, j), \ \textnormal{for} \ j \in S_J}%
$$
\end{minipage}
\begin{minipage}[m]{0.45\linewidth}
\small
$$
5. \quad \frac{j_1 \stackrel{\gamma}{\marktrans_J} j_2}%
		 	  {(i, j_1) \stackrel{\gamma}{\marktrans} (i, j_2), \ \textnormal{for} \ i \in S_I}%
$$
\end{minipage}

\end{definition}
\medskip

The first three clauses in Definition~\ref{def:composition} deal with interactive transitions: the first two tackle the independent evolution of each connector; the third one addresses their (synchronous) joint evolution.
Clauses 4 and 5 deal with Markovian transitions which are always  interleaved.

\begin{definition}[Synchronisation]\label{def:imcreo_synchronisation}
Let $I = (S_1\times S_2, Act, \intertrans, \marktrans, s)$ be an \imcreo model over a composite state space, and $M \subseteq \nodeset$. The synchronisation of $I$  with respect to $M$ is given by 
 $$\partial_M I = (S_M, Act\setminus M, \intertrans_M, \marktrans_M, s)$$
where  $S_M = \{ (i,j)\restricts_M \mid (i,j)\in S_1\times S_2\} $ and $ \intertrans_M$ and $\marktrans_M$ are the smallest relations satisfying, respectively, conditions 1 and 2 below:

\begin{minipage}[m]{0.45\linewidth}
\small
$$
1. \quad \frac{(i,j) \stackrel X \intertrans (i',j')  \quad (i,j) \nactblock M}%
		 {(i,j)\restricts_M \stackrel {X\setminus M} \intertrans_M (i',j')\restricts_M}%
$$
\end{minipage}
\begin{minipage}[m]{0.45\linewidth}
\small
$$
2. \quad \frac{(i,j) \stackrel \gamma \marktrans (i',j') \quad (R_{i'} \cup R_{j'}) \cap M = \emptyset }%
		 	  {(i,j)\restricts_M \stackrel {\gamma} \marktrans_M (i',j')\restricts_M}%
$$
\end{minipage}
\end{definition}
\end{document}